\RequirePackage{lineno}
\documentclass{article}

\usepackage[utf8]{inputenc}
\usepackage[T1]{fontenc}
\usepackage{authblk}
\usepackage{esvect}
\usepackage{musicography}

\usepackage{setspace}
\usepackage{amssymb}
\usepackage{times}
\usepackage{graphicx}
\usepackage{enumerate}
\usepackage{amsmath}
\usepackage{bm}
\usepackage{eucal}
\usepackage{url}
\usepackage{xcolor}
\usepackage{calc}
\usepackage{array}
\usepackage{indentfirst}
\usepackage{fancyhdr}

\usepackage[a4paper, total={6in, 8in}]{geometry}

\begin{document}
	
\title{Piano Timbre Development Analysis using Machine Learning}

\author{Niko Plath, Rolf Bader}
\affil{ Institute of Musicology\\ University of
	Hamburg\\ Neue Rabenstr. 13, 20354 Hamburg, Germany\\
}
\date{\today}

	%\setpagewiselinenumbers
	%\modulolinenumbers[1]
	%\linenumbers
	\doublespacing

	\maketitle

\begin{abstract}

A data set of recorded single played tones of a concert grand piano is 
investigated using Machine Learning (ML) on psychoacoustic timbre features. 
The examined instrument has been recorded at two stages: firstly right after manufacture and secondly after being played in a concert hall for one 
year. A previous study \cite{Plath2019} revealed that listeners clearly 
distinguished both stages but no clear correlation with acoustics, signal 
processing tools or verbalizations of perceived differences could be found. 
Using a Self-Organizing Map (SOM), training single as well as double feature 
sets, it can be shown that spectral flux is able to perfectly cluster the two 
stages. Sound Pressure Level (SPL), roughness, and fractal correlation 
dimension (as a measure for initial transient chaoticity) are furthermore able 
to order the keys with respect to high and low notes. Combining spectral flux 
with the three other features in double-feature training sets maintains stage 
clustering only for SPL and fractal dimension, showing sub-clusters for both 
stages. These sub-clusters point to a homogenization of SPL for stage 2 with 
respect to stage 1 and a pronounced ordering and sub-clustering of key regions 
with respect to initial transient chaoticity.
\end{abstract}

\section{Introduction}
In a previous study \cite{Plath2019}, the acoustics, timbre, and perception of a piano have been 
investigated at two stages, firstly right after manufacture  (further denoted as 
\textit{stage 1}) and secondly after one year of performance in a concert hall 
(\textit{stage 2}). 
In an ABX double-blind comparison test \cite{Clark1982} all 88 keys of both 
stages were presented to subjects. Additionally, subjects were asked to 
verbalize how they where able to distinguish between stages (regardless of whether they could tell correctly). The text input was cleaned and filtered utilizing natural language processing (NLP) methods and further categorized  
as affiliated to the domains  \textit{timbre}, \textit{pitch}, \textit{temporal}, \textit{spatial}, and \textit{loudness}.

Although listeners were significantly able to distinguish between the two stages, data analysis 
was not able to clearly identify the reason(s). Analyzed acoustical features 
 were driving point mobility, which did not change considerably, 
the crown bearing, and the action speed. Although the latter two showed 
differences, like a clear homogenization of the action speed over the year, they could not be 
associated with perception. 

Signal processing of the data sets showed deviations between the two stages: 
The spectral centroid was considerably lower for the mid- and high-frequency 
range one year after performance. Still this did not correlate with the distribution 
of verbalizations in the timbre category. Regarding pitch, the majority of 
verbalizations was given in frequency-ranges where no measurable pitch 
differences were present. Also, although verbalizations of temporal differences 
were reported, no considerable differences of attack time or decay time (RT60) could 
be observed. Regarding loudness domain features, Sound Pressure Level (SPL) and 
Interlevel Differences (ILD) decreased in the mid- and high-frequency range one 
year after playing, still verbalizations were too few to find correlations.

In this paper, a new approach is pursued to distinguish 
the two piano stages using Music Information Retrieval (MIR) parameters and 
Machine Learning (ML), namely a Self-Organizing Kohonen Map (SOM). 

\section{Method}

The \textit{apollon} and \textit{COMSAR} frameworks \cite{apollon,COMSAR}, 
developed at the Institute of Systematic Musicology over the last years, have 
been used within the project of \textit{Computational Phonogram Archiving }
\cite{Bader2019, 
Bader2021a, Bader2021b, Bader2021c}. The system uses the following timbre features to generate a feature vector:
spectral centroid, spectral spread, spectral flux, roughness, 
sharpness, SPL, and fractal correlation dimension. The fractal correlation 
dimension counts the amount of inharmonic partials and strong amplitude 
fluctuations which appear during the initial transient of musical 
sounds \cite{Bader2013}. Therefore it is a measure of the chaoticity of initial 
transients.

All 2$\times$88 single played tones are analyzed with respect to these features 
in frames of 50 ms over the cause of each sound and the arithmetic mean for each timbre 
feature is calculated. 
The aim is to cluster the two piano stages on a two-dimensional Kohonen map. 
Therefore, different feature combinations are used. At first, all features are 
trained by different maps as single features. Therefore the feature vector on 
each nodal point has only one entry. Secondly, different combinations of 
feature vectors are used and the resulting trained maps are selected according 
to clustering of piano stages as well as other sub-clusters within the two 
piano stages. It turns out that only single and double-features succeed in 
finding such clusters.

\section{Results}

\subsection{Single feature SOMs}

\begin{figure}[!]
	\begin{center}
		\includegraphics[width=1\columnwidth]{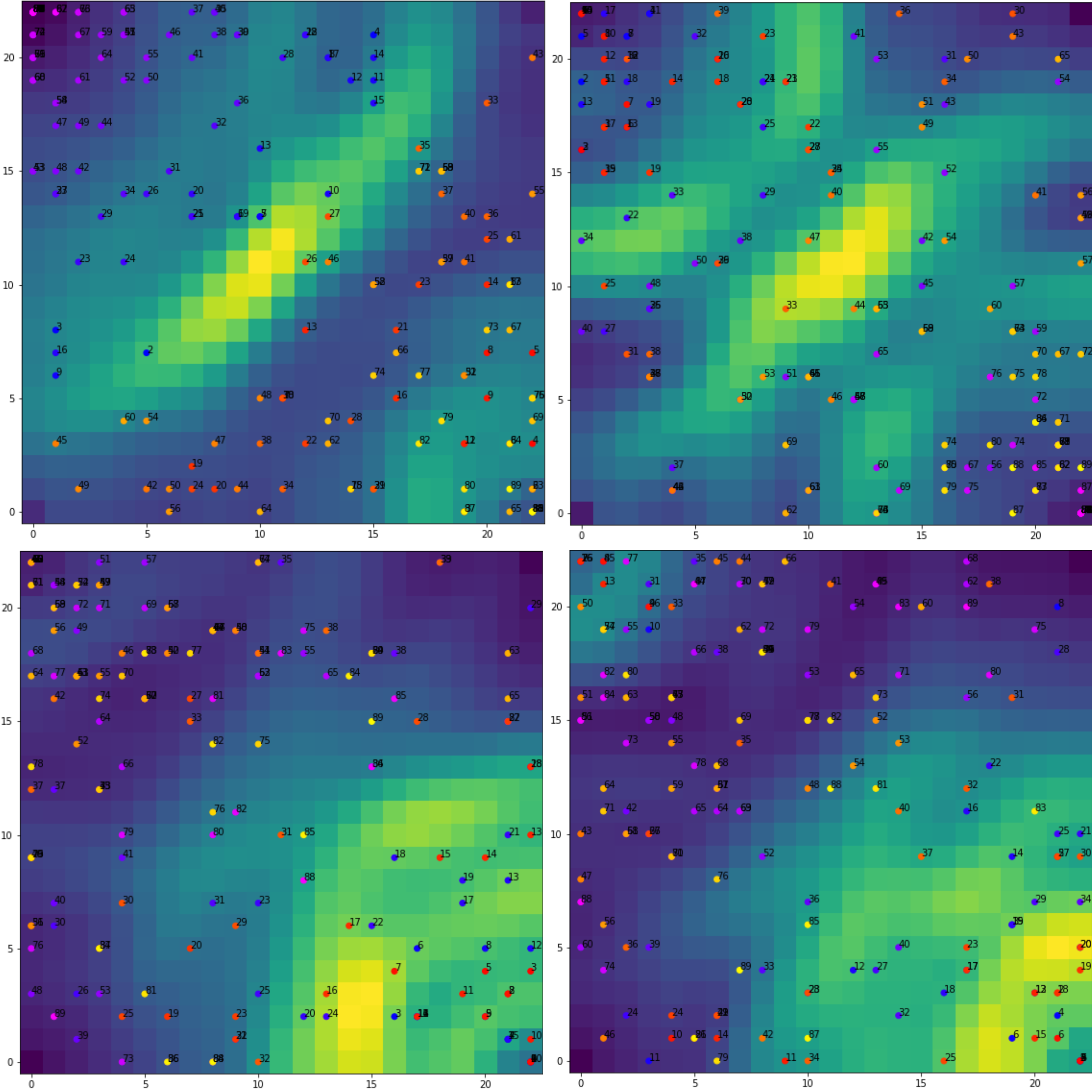}
		\includegraphics[width=.7\columnwidth]{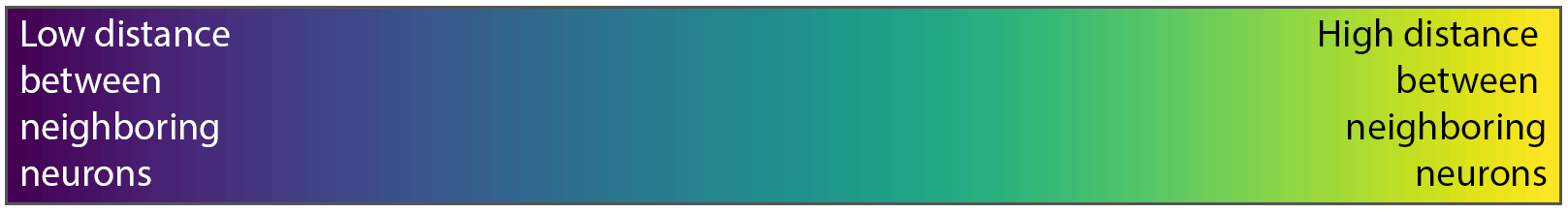}
		
		\caption{SOM for four single timbre features: spectral flux (upper left), SPL (upper right), roughness (lower left), and fractal correlation dimension (lower right). Stage 1 pianos keys: red(lower keys)-yellow(upper keys), stage 2 piano keys: blue(lower keys)-violet(upper keys). Background: u-matrix displaying similarity between neighboring neurons. Spectral flux clusters the pianos perfectly. SPL, roughness, and fractal correlation cluster according to keys.}
	\end{center}
\label{SOM_Single}
\end{figure}

In Fig. \ref{SOM_Single} four single timbre features are used to train the SOM. 
For the two-dimensional maps tones of all 88 keys at two stages are displayed as 
dots, where the hue is changed from lowest to highest keys. Stage 1 keys are 
red (lowest) to yellow (highest), stage 2 are blue (lowest) to violet 
(highest). Additionally, the key number is shown next to the dot. Sometimes 
more than one key is placed on one neuron. The background color shows the 
u-matrix of the SOM, which is the distance between neighboring neurons. Darker 
colors imply high similarity for neighboring neurons, lighter colors 
like yellow mean that the neighboring neurons are very different. 
Therefore, yellow ridges indicate boundaries between more homogeneous regions 
displayed in blue.

Spectral flux (upper left) is able to perfectly cluster the two piano stages, where stage 1 is on the upper right and stage 2 on the lower left side of the map. Both stages are clearly separated by a strong ridge in the u-matrix (yellow line). The SOM orders spectral flux as a continuous decrease from a maximum at the lower right corner to a minimum at the upper left corner. Therefore, stage 1 has considerably more flux than stage 2. This might be the main reason for the successful separation of the two stages by listeners found in \cite{Plath2019}. Furthermore, stage 2 shows a continues decay of flux from lower keys to higher ones, where stage 1 is not at all so clearly sorted.

SPL (upper right) is not able to cluster the stages, but clearly puts the low keys on the upper left corner where SPL is high and the high keys in the lower right corner where SPL is low. Additionally, next to a main ridge separating the lower right from the upper left corner, two ridges separate a middle region. Therefore, in terms of SPL, four clusters are found, where the two middle clusters sort keys in the medium-frequency range.

Roughness (lower left) clusters lower keys of both stages at the lower left corner and distributes all other keys in the rest of the map. This region of strong roughness, expected from lower keys is strong in the u-matrix (yellow region) meaning that additionally to strong roughness, the low keys are also very diverse with respect to roughness.

The map for fractal correlation dimension (lower right) behaves similarly to the one for roughness, pointing to a correlation between the two parameters with respect to piano sounds. Training a SOM with only these two timbre features does not result in a perfect orthonormal relation, but also not in a perfect correlation.

Therefore, spectral flux is assumed to be the main cause of listeners being able to distinguish the two stages aurally. The other timbre parameters go into details which is further investigated using timbre combinations below. Other timbre features did not result in any clear clusters.

\subsection{Double-feature SOMs}

As shown with multi-dimensional scaling (MDS) listening tests over the last decades, timbre perception is multi-dimensional (for a review see \cite{Bader2013}). Therefore it is likely that listeners use a combination of timbre features for distinguishing between tones. Almost all MDS investigations resulted in a maximum of three timbre features, where the third one is most often very weakly present, hereinafter combinations of two features are presented. Since spectral flux is the only feature being able to distinguish between stage 1 and 2 and as listeners were asked for performing a difference discrimination, spectral flux was used in all three combinations discussed below. The respective second features used in combination with spectral flux are SPL, roughness, and fractal correlation dimension, as again only these features did result in reasonable clusters as single features (see above).

\begin{figure}[!]
	\begin{center}
		\includegraphics[width=1\columnwidth]{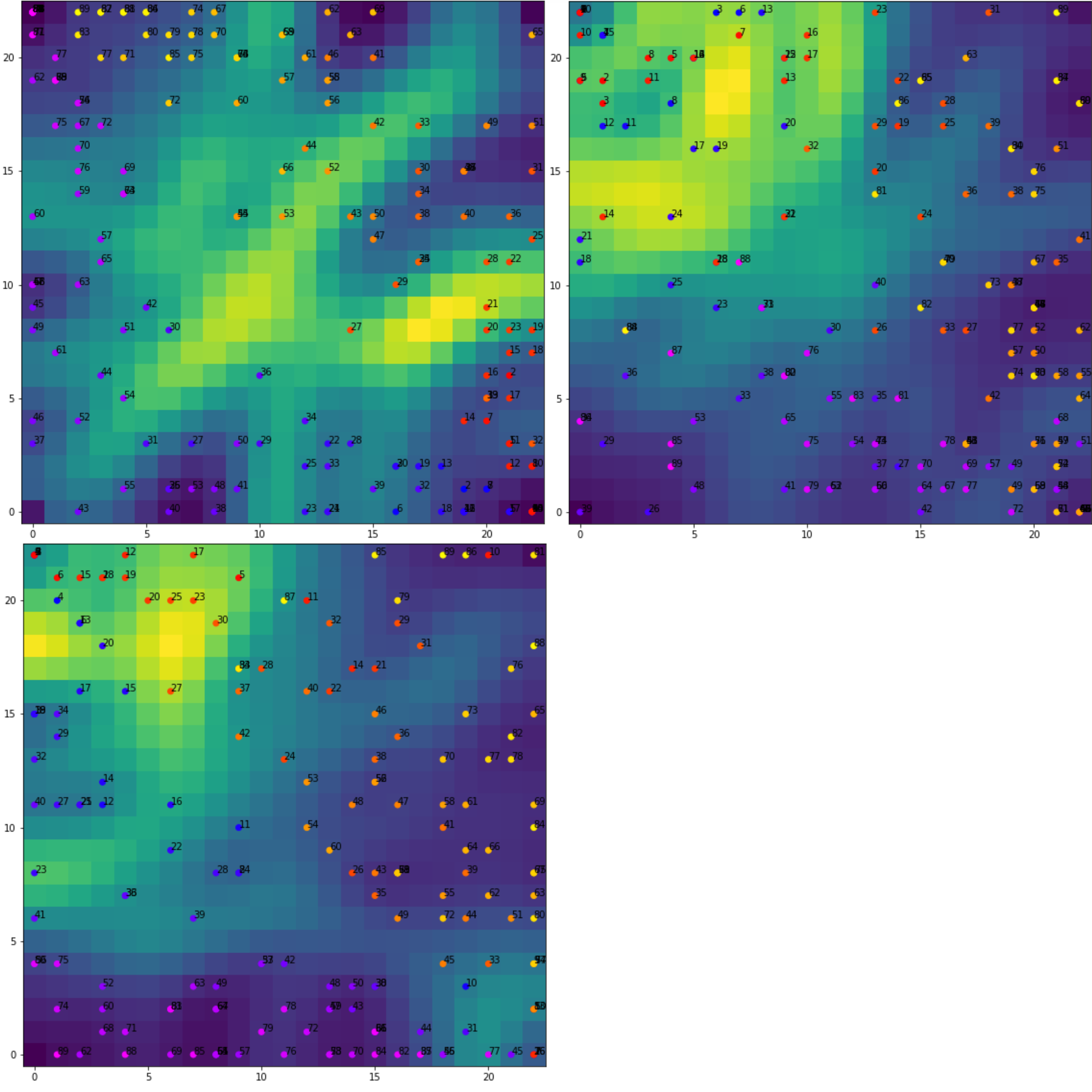}\\
		\includegraphics[width=.7\columnwidth]{viridis_legend}
		\caption{SOM for three double-feature correlations: spectral flux/SPL (upper left), spectral flux/roughness (upper right), and spectral flux/fractal correlation dimension (lower left). Coloring like in \ref{SOM_Single}. While the combinations with SPL and fractal correlation maintain perfect clustering of stage 1 and 2, combining with roughness mixes the clusters. Also additional sub-clustering is observable.}
	\end{center}
	\label{SOM_Double}
\end{figure}

\begin{figure}
	\centering
	\includegraphics[width=1\linewidth]{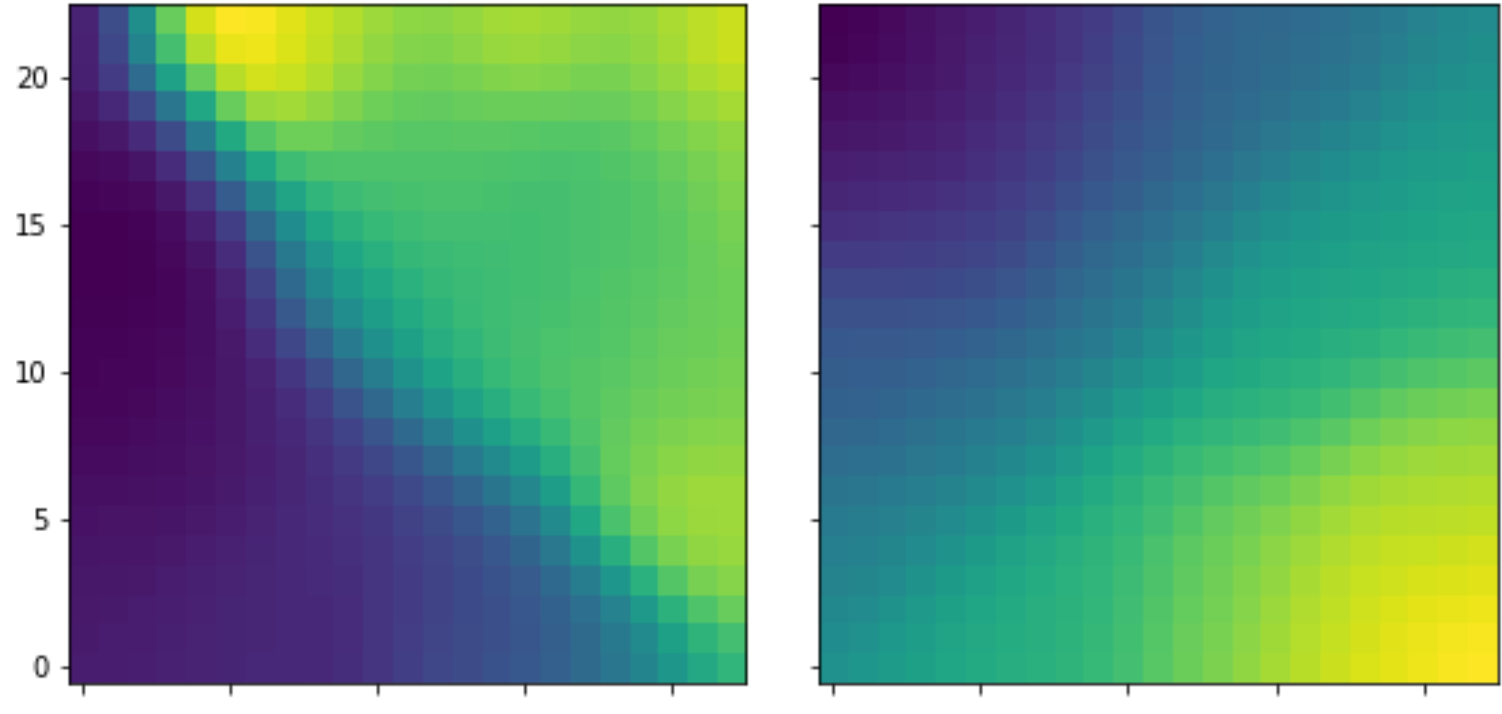}
	\caption{Feature distribution on SOM for the combination spectral flux/SPL of Fig. \ref{SOM_Double} (upper left). Left: spectral flux, right: SPL. Yellow: large values, blue: low values. The two features are nearly perfectly orthonormal on to another.}
	\label{fig:pianokicomponentleftfluxrightspl}
\end{figure}

The combination of spectral flux vs. SPL in Fig. \ref{SOM_Double} (upper left) maintains a perfect cluster of the two stages, where stage 1 is in the upper right and stage 2 in the lower left corner. Additionally, both stages are clearly ordered in terms of their keys, where lower keys are in the lower right and upper keys in the upper left corner. The distribution of the two features over the map are almost perfectly orthonormal, as shown in Fig. \ref{fig:pianokicomponentleftfluxrightspl}. Again, the spectral flux is stronger for stage 1, and SPL is stronger for lower keys. Additionally, a distinct ridge splits the lower keys from the upper ones, going from the lower left to the upper right corner. Another strong ridge can be seen in the lower right side which is considerably stronger for stage 1 than for stage 2. Another smaller ridge is found in the upper left corner, again a bit stronger for stage 1 than for stage 2. These ridges clearly separate stage 1 into four distinct regions, ordered in terms of low keys, lower middle keys, upper middle keys, and high keys. Such a separation can also slightly be seen for stage 2, although it is less clearly developed. This points to a homogenization of SPL from stage 1 to stage 2, without influence on spectral flux.

\begin{figure}
	\centering
	\includegraphics[width=1\linewidth]{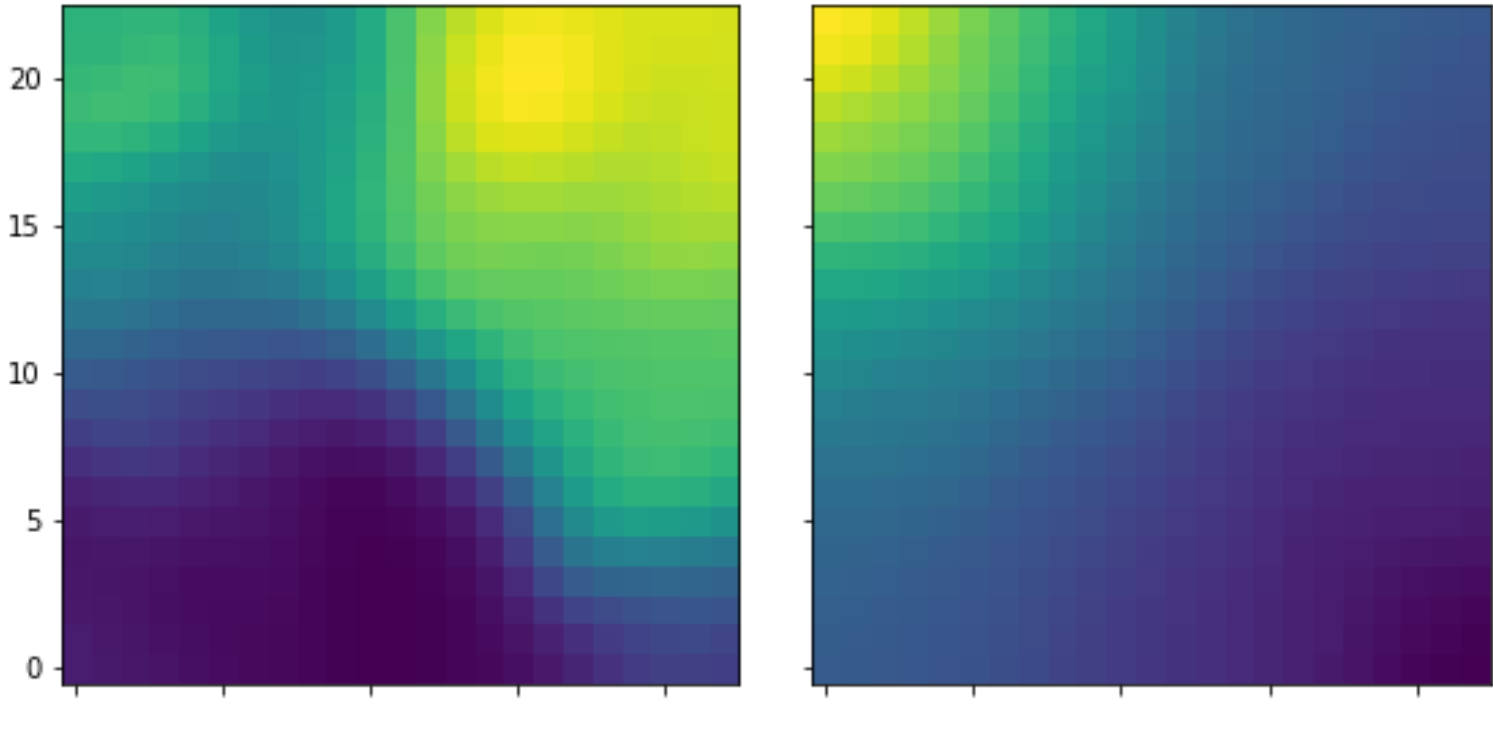}
	\caption{Distribution of features spectral flux (left) and roughness (right) for SOM shown in Fig. \ref{SOM_Double}, upper right. Color as in Fig. \ref{fig:pianokicomponentleftfluxrightspl}.}
	\label{fig:pianokicomponentleftfluxrightroughness}
	\end{figure}

The combination of spectral flux vs. roughness, shown in Fig. \ref{SOM_Double} (upper right) is no longer able to perfectly cluster the stages. The feature distribution, shown in Fig. \ref{fig:pianokicomponentleftfluxrightroughness} is also not perfectly orthonormal, flux is strongest at the upper right side, as expected, is weakened towards to upper left side only to get stronger again at the upper left corner. This is reflected in the distribution of stage 1 keys of both pianos at the upper left corner. The roughness distribution is clearer, with a maximum at the upper left corner, where the low keys are located, smoothly decreasing towards the lower right corner. Due to the decrease in clustering of stage 1 and stage 2 this combination of timbre features seems unlikely to be used by listeners.

\begin{figure}
	\centering
	\includegraphics[width=1\linewidth]{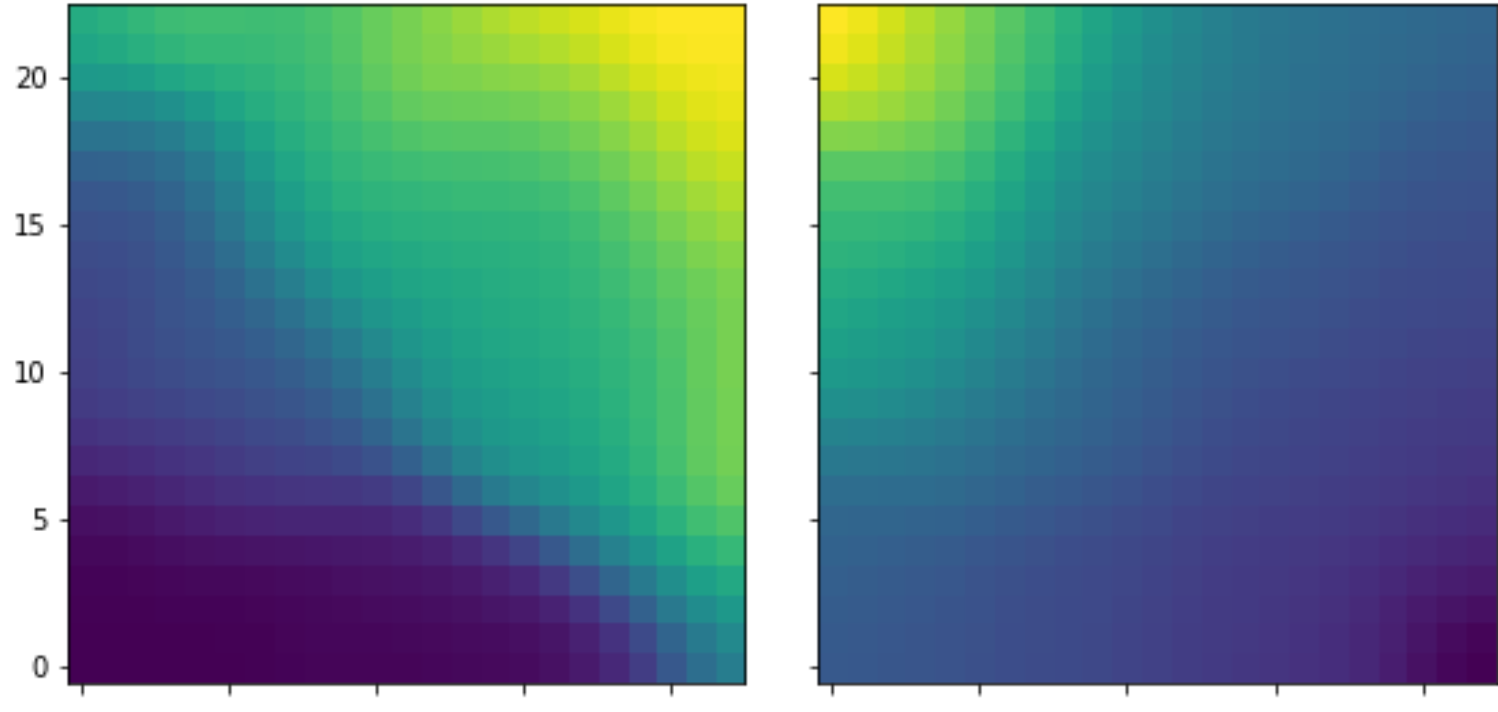}
	\caption{Distribution of features spectral flux (left) and fractal correlation dimension (right) for SOM shown in Fig. \ref{SOM_Double}, lower left. Color as in Fig. \ref{fig:pianokicomponentleftfluxrightspl}.}
	\label{fig:pianokicomponentleftfluxrightfractal}
\end{figure}

The last combination shown is spectral flux vs. fractal correlation dimension in Fig. \ref{SOM_Double}, lower left. The feature distributions shown in Fig. \ref{fig:pianokicomponentleftfluxrightfractal} are now again nearly perfectly orthonormal to one another, with flux being strongest in the upper right corner, where stage 1 is located, and fractal dimension being strongest in the upper left corner, where low keys are distributed. Interestingly, stage 2 pianos have a sub-cluster at the lower end of the plot, separated from upper mid-frequency keys by a ridge. Another sub-cluster is found at the left side roughly between neurons 11-16. Stage 1 sounds are sub-clustered in a different way, where two sub-clusters are separated by a ridge from the upper right corner to the center of the plot. Additionally, stage 1 keys 87 and 83 (yellow) are found around neurons (10-12, 17-20), so in the range of low keys. Overall, stage 2 pianos have a much clearer sub-clustering and smoother key distribution, pointing to a more systematic fractal dimension distribution. As the fractal dimension displays the chaoticity in the initial transient of the signal, the transient of stage 2 seems to be clearer organized.

\section{Conclusions}

Spectral flux clusters the two piano stages perfectly and therefore is assumed to be the main cause of the clear separation of the two stages by listeners found in \cite{Plath2019}. Furthermore, stage 2 has a much  smoother distribution of SPL and a clearer sub-clustering structure of the chaoticity of initial transients. This is expected due to the one-year treatment of piano tuners of stage 2 compared to state 1, which seem to have these features as their aim. Assuming at least two timbre parameters to be used by listeners to separate the two stages, two candidates appear from the present analysis, being spectral flux with SPL and spectral flux with fractal correlation dimension.

	\newpage
\addtocounter{page}{2}

\end{document}